\documentclass[12pt, a4paper]{article}
\usepackage[font=footnotesize]{caption}
\usepackage[utf8]{inputenc}
\usepackage{graphicx}
\usepackage{authblk}
\usepackage{booktabs}
\usepackage{verbatim}
\usepackage{url}
\usepackage{caption}
\usepackage{subcaption}
\usepackage{indentfirst}
\usepackage{url}
\usepackage{amssymb}
\usepackage{amsmath}
\usepackage{multirow}
\title{Random Walks Performed by \\ Topologically-Specific Agents on \\ Complex Networks}
\author{Alexandre Benatti$^1$ and Luciano da F. Costa$^2$}

\affil{
$^1$Institute of Mathematics and Statistics - DCC \\
University of S\~ao Paulo \\
Rua do Mat\~ao, 1010, S\~ao Paulo, SP 05508-090 Brazil %\\ \vspace{0.5cm}
\\ \vspace{0.5cm}
$^2$S\~ao Carlos Institute of Physics - DFCM \\
University of S\~ao Paulo \\
Av. Trabalhador S\~ao-Carlense, 400, \\
S\~ao Carlos, SP 13566-590 Brazil \\
(\emph{Prof. Senior})
}

\date{current version 15th May. 2025}

\begin{document}

\maketitle

\begin{abstract}
Random walks by single-node agents have been systematically conducted on various types of complex networks in order to investigate how their topologies can affect the dynamics of the agents. However, by fitting any network node, these agents do not engage in topological interactions with the network. In the present work, we describe random walks on complex networks performed by agents that are actually small graphs. These agents can only occupy admissible portions of the network onto which they fit topologically, hence their name being taken as topologically-specific agents. These agents are also allowed to move to adjacent subgraphs in the network, which have each node adjacent to a distinct original respective node of the agent. Given a network and a specific agent, it is possible to obtain a respective associated network, in which each node corresponds to a possible instance of the agent and the edges indicate adjacent positions. Associated networks are obtained and studied respectively to three types of topologically-specific agents (triangle, square, and slashed square) considering three types of complex networks (geometrical, Erd\H{o}s-R\'enyi, and Barab\'asi-Albert). Uniform random walks are also performed on these structures, as well as networks respectively obtained by removing the five nodes with the highest degree, and studied in terms of the number of covered nodes along the walks. Several results are reported and discussed, including the fact that substantially distinct associated networks can be obtained for each of the three considered agents and for varying average node degrees. Respectively to the coverage of the networks by uniform random walks, the square agent led to the most effective coverage of the nodes, followed by the triangle and slashed square agents. In addition, the geometric network turned out to be less effectively covered.
\end{abstract}

\section{Introduction}

The areas of random walks
(e.g.~\cite{berg1993random,shlesinger1995levy,lawler2010random,revesz2013random,ibe2013elements,spitzer2013principles,xia2019random}) and complex networks 
(e.g.~\cite{albert2002statistical, costa2007characterization,costa2011analyzing,newman2018networks}) have motivated substantial interest from the scientific community, which has taken place both independently as well as in an integrated manner (e.g.~\cite{noh2004random,costa2006visual,da2007exploring,brockmann2008anomalous,rosvall2008maps,fronczak2009biased,lambiotte2015effect,de2017knowledge,masuda2017random}).

Given that complex networks often present an intricate topological (and sometimes also geometrical) structure, they provide a particularly interesting environment in which to perform random walks. Interest is often focused on studying how specific topological (and geometrical) properties of complex networks can influence or even determine the dynamic properties of the respectively performed random walks (e.g.~\cite{noh2004random, fortunato2007random,da2007correlations,travenccolo2009border,viana2012effective,de2019connecting,regnier2023universal}).

However, despite the continuing interest in studying several types of random walks on complex networks, most of the reported approaches consider that the moving agent(s) are point-like structures, occupying a single network node at each step.  When this type of agent performs random walks on complex networks, the respective movements can be influenced only by the topology of the network. However, in case the agent consisted of a small graph, its motion would be influenced not only by the topology of the network, but also by its specific topological properties.

\begin{figure}[!ht]
  \centering
     \includegraphics[width=.6 \textwidth]{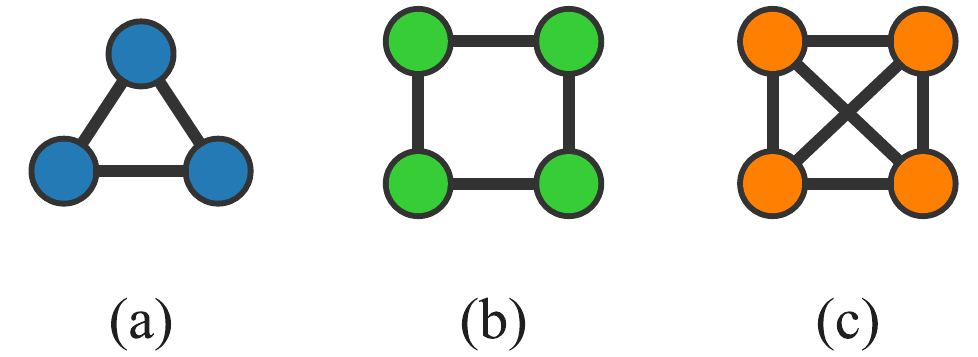}
   \caption{Three examples of topologically-specific moving agents: (a) triangle, (b) square, and (c) slashed square, consisting of relatively small graphs, henceforth considered for performing random walks in complex networks. These agents are allowed only to occupy a subgraph of the network provided their topology is completely contained in that subgraph. An agent can then move to new adjacent subgraphs of the networks where it fits, therefore performing a respective random walk.}\label{fig:agents}
\end{figure}

The present work aims at developing and studying concepts and methods allowing random walks on complex networks performed by relatively more complex agents -- henceforth referred to as being \emph{topologically-specific} --- corresponding to relatively small connected graphs, such as those illustrated in Figure~\ref{fig:agents}.

The random walks performed by the types of agents illustrated in Figure~\ref{fig:agents} are henceforth said to be topologically-specific because they are only allowed to occupy portions of a given network corresponding to subgraphs that contain the respective agent. Furthermore, when at a specific place in the network, a topologically-specific agent can only move to adjacent positions into which they also fit, as illustrated in Figure~\ref{fig:moviment}.

Several are the reasons for defining and studying random walks performed by topologically-specific moving agents in complex networks. First and foremost, is the fact that the type of subgraph defining the agent will have a critical influence on the respective displacements respectively to different types of complex networks. For instance, the agent in Figure~\ref{fig:agents}(b) will tend to have enhanced mobility in complex networks resembling orthogonal lattices. Interestingly, the motion of the topologically-specific agents then becomes closely dependent not only on the respective network (as is often the case with traditional random walks), but also on how their own specific topology relates and matches the topology of the network onto which the random walks are to be performed.

The present work is aimed at describing the basic concepts and methods underlying a possible approach to topologically-specific moving agents and then to study the coverage (e.g.~\cite{chupeau2015cover, maier2017cover}) and specific types of displacements of these agents while considering three main types of complex networks: (i) ER -- Erd\H{o}s-R\'enyi~\cite{erdos1959random}; (ii) BA -- Barab\'asi-Albert~\cite{barabasi1999emergence}; (iii) GEO -- geometric (Delaunay)~\cite{riedinger1988delaunay}.

Uniform random walks are considered, corresponding to choosing among new positions. Several interesting results are reported and discussed. The displacement of the topologically-specific agents after a single step of the random walks was also studied in terms of the respective number of moving nodes, also with interesting results, especially concerning the BA networks.  An additional experiment is also reported in which the $5\%$ of nodes with the highest degrees (henceforth understood as \emph{hubs}) were removed from the network.  This situation has been considered as a means for identifying the possible effect of network hubs on the overall node coverage efficiency resulting from the considered types of nodes while moving on the three adopted network models.

The present work initiates by describing the adopted concepts and methods for the performance of topologically-specific random walks on complex networks, as well as the concept of the associated network. Next, experimental results concerning the structural analyses of the associated networks obtained from the ER and BA models, including examples of topologically-specific moving agents and the node coverage by uniformly random walks, are presented and discussed.

\section{Basic Concepts}\label{sec:methods}

In this section, we introduce the main basic concepts and methods used in our study. 

\subsection{Network Models}\label{sec:models}

Three types of \emph{network topologies} have been considered in this work. Regular Erd\H{o}s-R\'enyi (ER) structure~\cite{erdos1959random}, Barab\'asi-Albert (BA) networks~\cite{barabasi1999emergence}, and geometric network (GEO), generated through the implementation of the Waxman method~\cite{waxman1988routing}, have also been considered. Henceforth, all networks are assumed to have a single connected component.

These three types have mostly distinct properties, therefore allowing a more general study of the topologically-specific random walks. Basically, the ER model consists of a network with uniformly random structure characterized by most nodes having similar degrees, while the BA structures are characterized by degree heterogeneity (power law).  Both these models tend to have relatively small average shortest paths. Contrariwise, the GEO networks tend to have large average shortest path, while presenting nodes with relatively homogeneous degrees.

\subsection{Topologically-Specific Agents}

A \emph{topologically-specific agent} is henceforth taken to correspond to a relatively small connected graph when compared to the size of the complex network to be explored. In this work we considered the agents shown in Figure~\ref{fig:agents}. For comparative purposes, a single point is also taken into account. The \emph{size} $n$ of an agent corresponds to the number of its nodes. The size of the complex network is $N \gg n$.

A given topologically specific agent is said to \emph{fit} or \emph{occupy} a specific locus of the complex network provided that locus corresponds to one of the network subgraphs that completely contains the graph representing the specific agent. In the present work, the fact a network locus contains the agent graph will be referred to as that locus being respectively \emph{admissible} to the agent.

\subsection{Movement of Agents}

The method adopted in the present work in order to choose a possible next position of an agent is explained as follows respectively to isomorphic agents.

Given that an agent of size $n$ is at a specific current locus of the complex network to be explored, the first-neighbors of each of its $n$ nodes are identified and organized as a respective list (or vector). Each original node is also incorporated into the same respective list, allowing the node to remain at the same position.

Then, a node is drawn in a uniformly random manner from each of the $n$ respective lists of respective first-neighbors described above. The obtained set of $n$ nodes is then used to extract the respective subgraph $\Gamma$ from the network containing those nodes. The topology of the agent is then compared to that of the subgraph $\Gamma$. In case the subgraph $\Gamma$ is found to \emph{contain} the agent, the latter is moved into its new locus. Two or more nodes are not allowed to proceed to a same destination node, therefore ensuring preservation of the subgraph size and topology (e.g.~a triangle agent remains so during the whole random walk).

\begin{figure}
  \centering
     \includegraphics[width=.8 \textwidth]{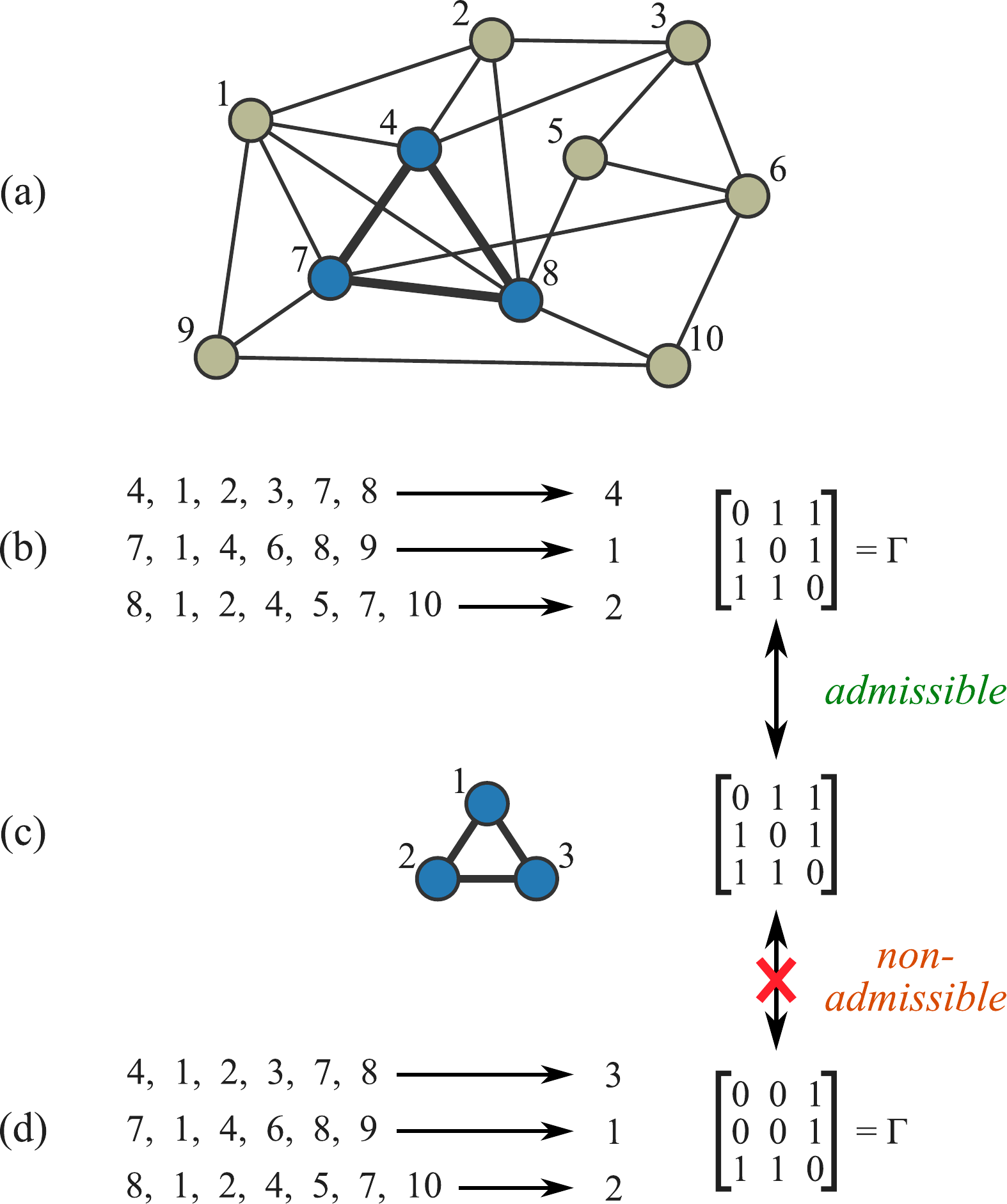}
   \caption{Illustration of the method adopted in this work for choosing a possible next move for a topologically-specific agent. The current position of the agent ($n=3$ in the network is shown in blue in (a). Three lists of first-neighbors are obtained for each of the agent nodes, including these nodes (b). Three node labels are respectively chosen from the lists with uniform probability, defining a respective subgraph from the larger network. The resulting adjacency matrix is compared to the adjacency matrix of the agent (c). In case this subgraph matrix contains the agent, as verified in (b), the agent can proceed to this possible new position, which is not the case in situation (d).}\label{fig:method}
\end{figure}

Figure~\ref{fig:method} illustrates the above methodology for identifying a possible subsequent (adjacent) new position for a topologically-specific moving agent corresponding to a triangle (with size $n=3$).

Figure~\ref{fig:method}(a) illustrates an agent occupying a specific locus of a larger embedding network. The first-neighbors of each of the three nodes of the agent are identified, giving rise to respective lists shown in (b), which also include the associated original nodes. Three labels are then chosen from each of these lists with uniform probability yielding, in the specific case of this figure, the labels $4, 1, 2$. The subgraph defined in the network by these labels is then identified, and its adjacency matrix obtained. In case the adjacency matrix of the network subgraph is found to contain the adjacency matrix of the topologically-specific agent (c), the locus $4, 1, 2$ is understood to be admissible to the agent, so that the latter can proceed to the former. Also illustrated in figure~\ref{fig:method}(d) is a situation in which, though adjacent to the current locus occupied by the agent, the locus $3, 1, 2$ is non-admissible, therefore precluding the respective agent displacement.

In case the considered agent is non-isomorphic, the above described comparison approach needs to be applied to all possible permutations.

\begin{figure}[!ht]
  \centering
     \includegraphics[width=.8 \textwidth]{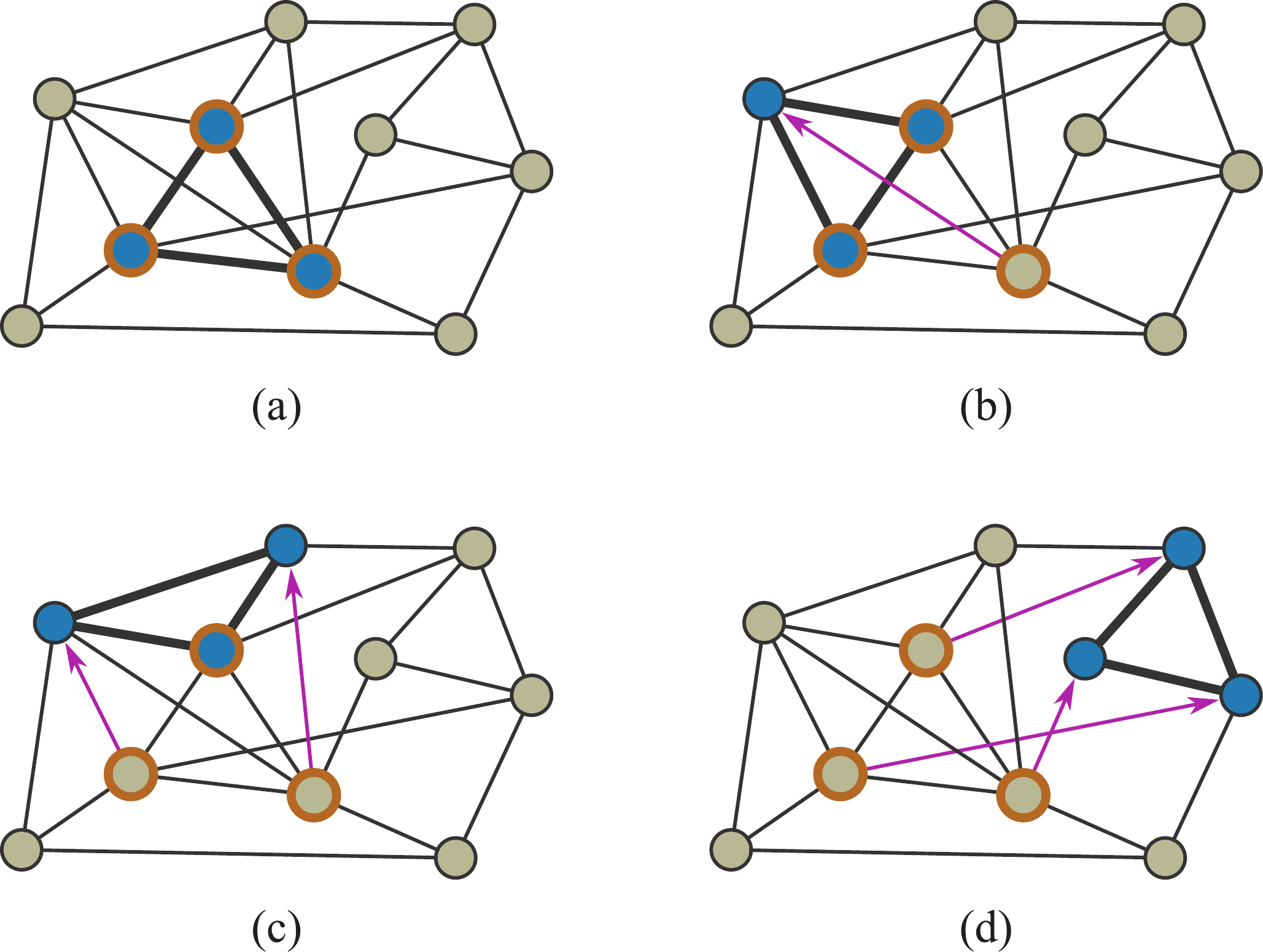}
   \caption{A triangular agent (in blue) contained in a network (a) and some of its possible displacements involving 1 (b), 2 (c), and 3 (d) nodes. The arrows (in magenta) indicate the motion of each of the agent nodes. 
 The nodes corresponding to the initial position of the agent are shown with orange borders.}\label{fig:moviment}
\end{figure}

In addition to the above-discussed critical relationship between the topology of the agent and a given complex network, Figure~\ref{fig:moviment} shows examples of how a topologically-specific agent can move at each subsequent step. While in more traditional random walks, the agent moves to the destination adjacent cell in a unique manner, there are several ways in which a topologically-specific agent can move from its current position. For instance, in Figure~\ref{fig:moviment}(b-d), the agent will have only one, two, or three of its nodes respectively displaced.

\subsection{Associated Networks}\label{sec:equivNet}

It follows from the above description that the displacement of a topologically-specific from a current locus to a new locus requires the two following conditions to be satisfied: (1) the new position corresponds to a subgraph that is \emph{adjacent} to the subgraph currently occupied by the agent, and that (2) the new position is \emph{admissible} to that specific agent.

\begin{figure}[!ht]
  \centering
     \includegraphics[width=.99 \textwidth]{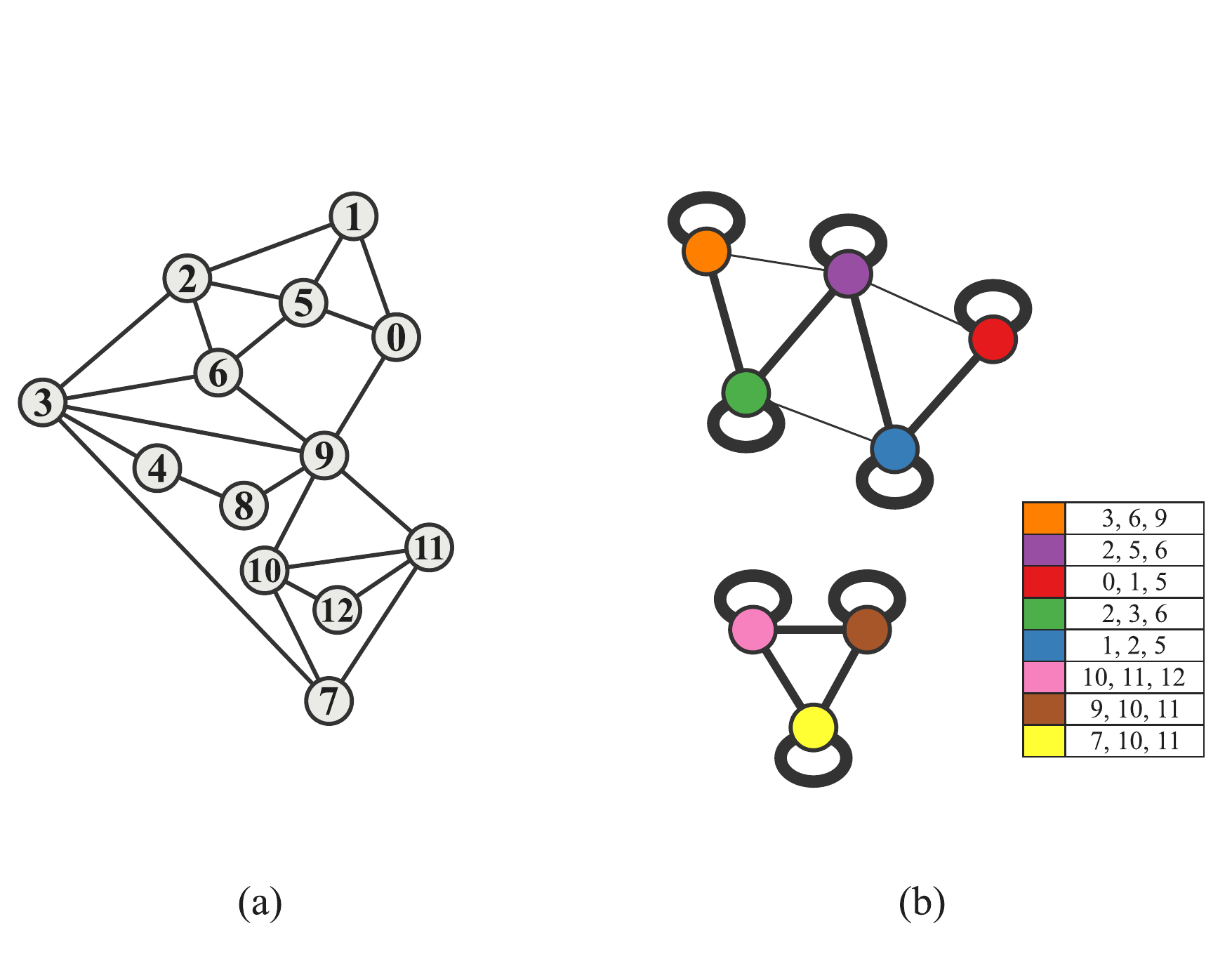}\\
   \caption{Example of a simple network (a) and a respective associated network (b). The original network nodes corresponding to each associated network node are indicated by colors as indicated in the legend. Note that the associated network is a weighted graph in which the weights of the edges are shown proportional to the transition probability of the agent.}\label{fig:ex_equivNet}
\end{figure}

This makes it possible to define a respectively \emph{associated network} that is derived from an initial complex network by considering an agent's walks through the network. Each node in the associated network is defined as a potential position of the agent within the original network (see Fig.~\ref{fig:ex_equivNet}). In contrast, the edges between these nodes represent the possibility of the agent moving from one position to another. Also, it should be observed that the associated network obtained for a given network respectively to a specific agent can have a number of nodes which can be larger, smaller, or equal to the size of the original network.

Given a network and an agent, the representation of the admissible positions and displacements of the latter in terms of an associated network paves the way to some interesting possibilities, including a more direct visualization of the just mentioned properties. For instance, the number of nodes in the associated network indicates the number of times the considered agent fits into the original network. On the other hand, the edges represent the possible paths which the agent can move between these positions, while the weights of the edges express the number of possible movements between these positions.

A further advantage of this approach is that performing a random walk in the associated network becomes computationally more efficient, as operating within an associated network simplifies the calculations and the exploration process compared to performing the movement of topologically-specific agents directly in the original network.

\begin{figure}[!ht]
  \centering
     \includegraphics[width=.9 \textwidth]{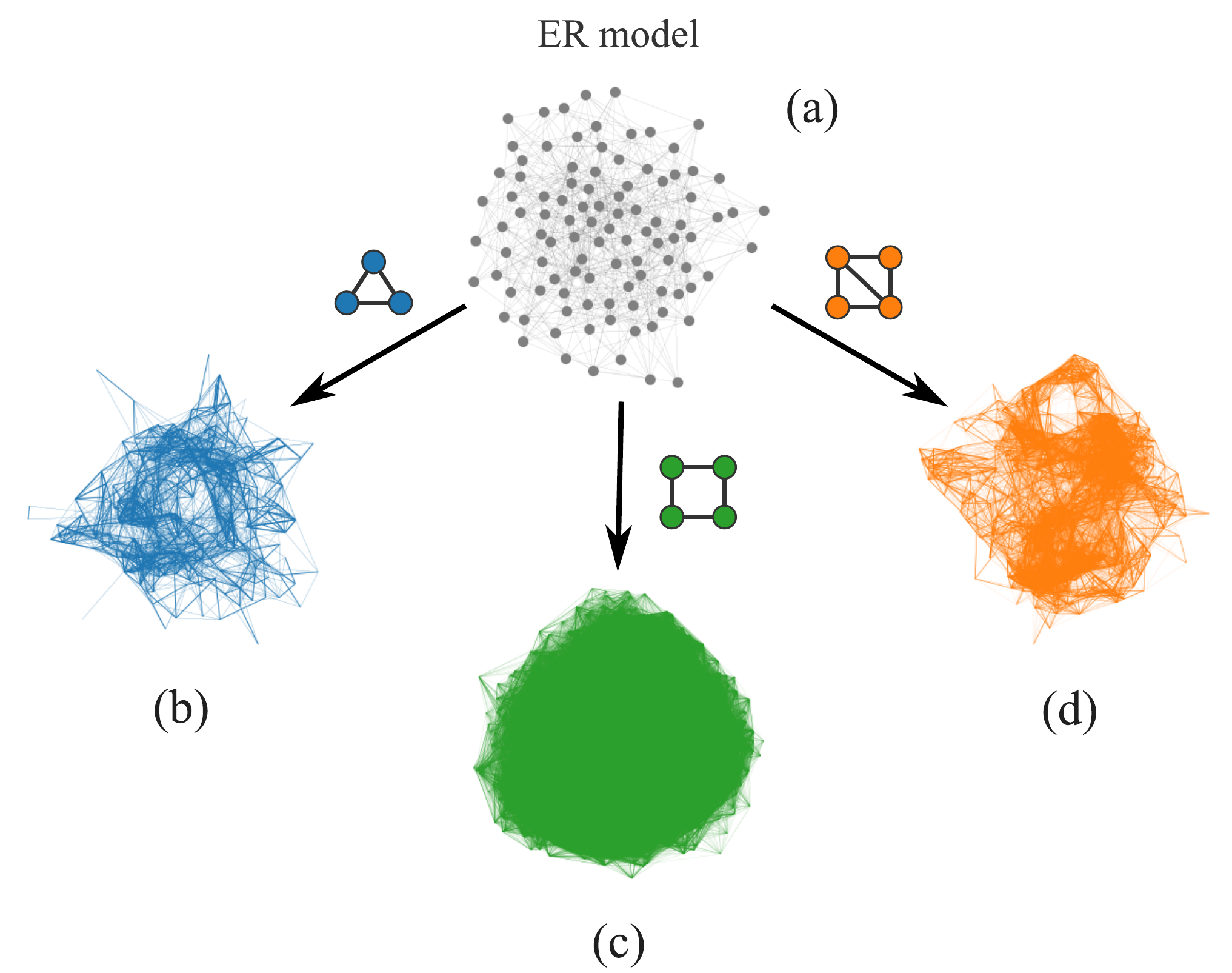}
   \caption{Illustration of an ER network (with 100 nodes and $<k=12>$) (a) and the respective associated networks obtained for the triangle (b), square (c), and slashed square (d) agents.}\label{fig:net2equiv_ER}
\end{figure}

\begin{figure}[!ht]
  \centering
     \includegraphics[width=.9 \textwidth]{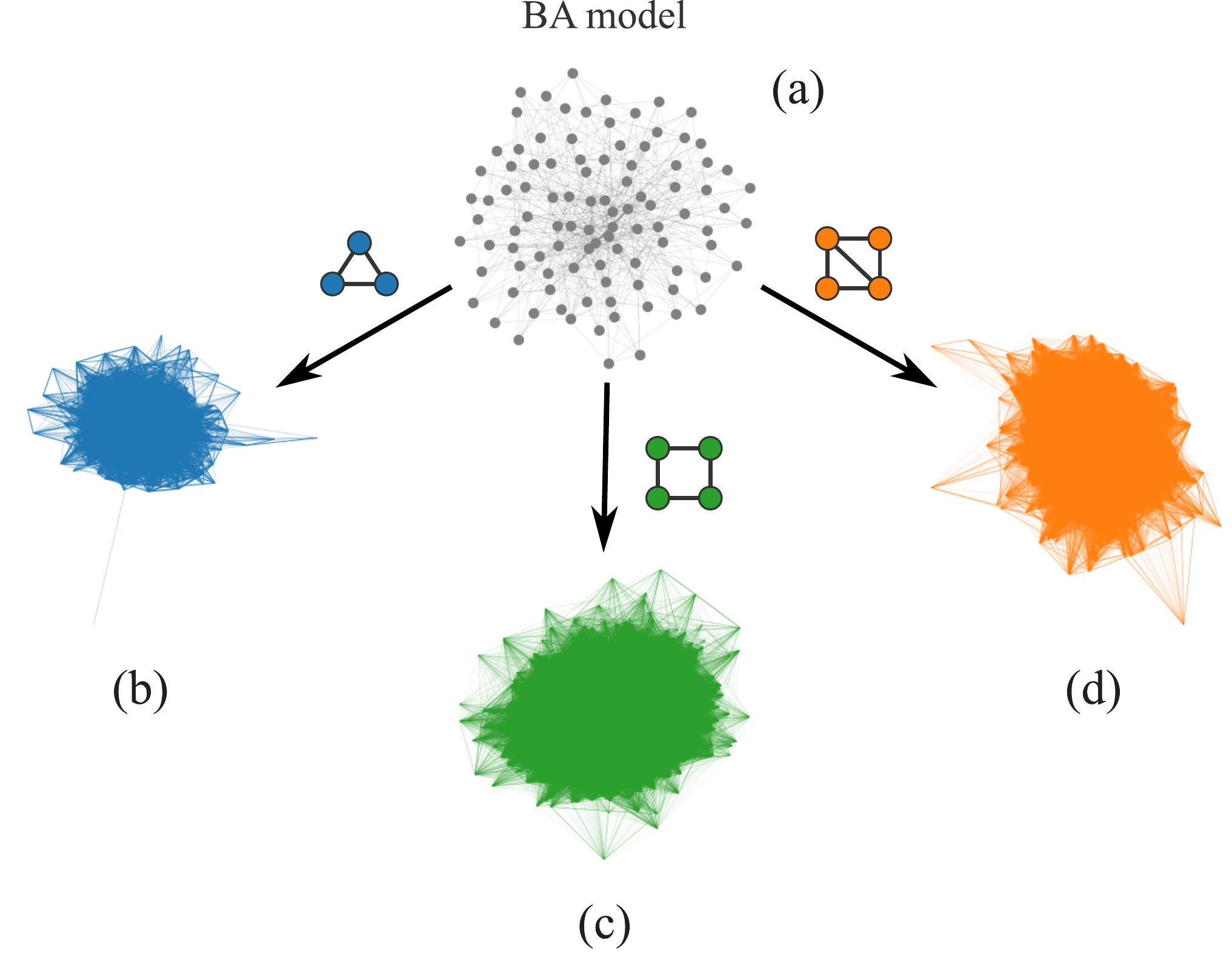}
   \caption{Illustration of a BA network (with 100 nodes and $<k=11.58>$) (a) and the respective associated networks obtained for the triangle (b), square (c), and slashed square (d) agents.}\label{fig:net2equiv_BA}
\end{figure}

\begin{figure}[!ht]
  \centering
     \includegraphics[width=.9 \textwidth]{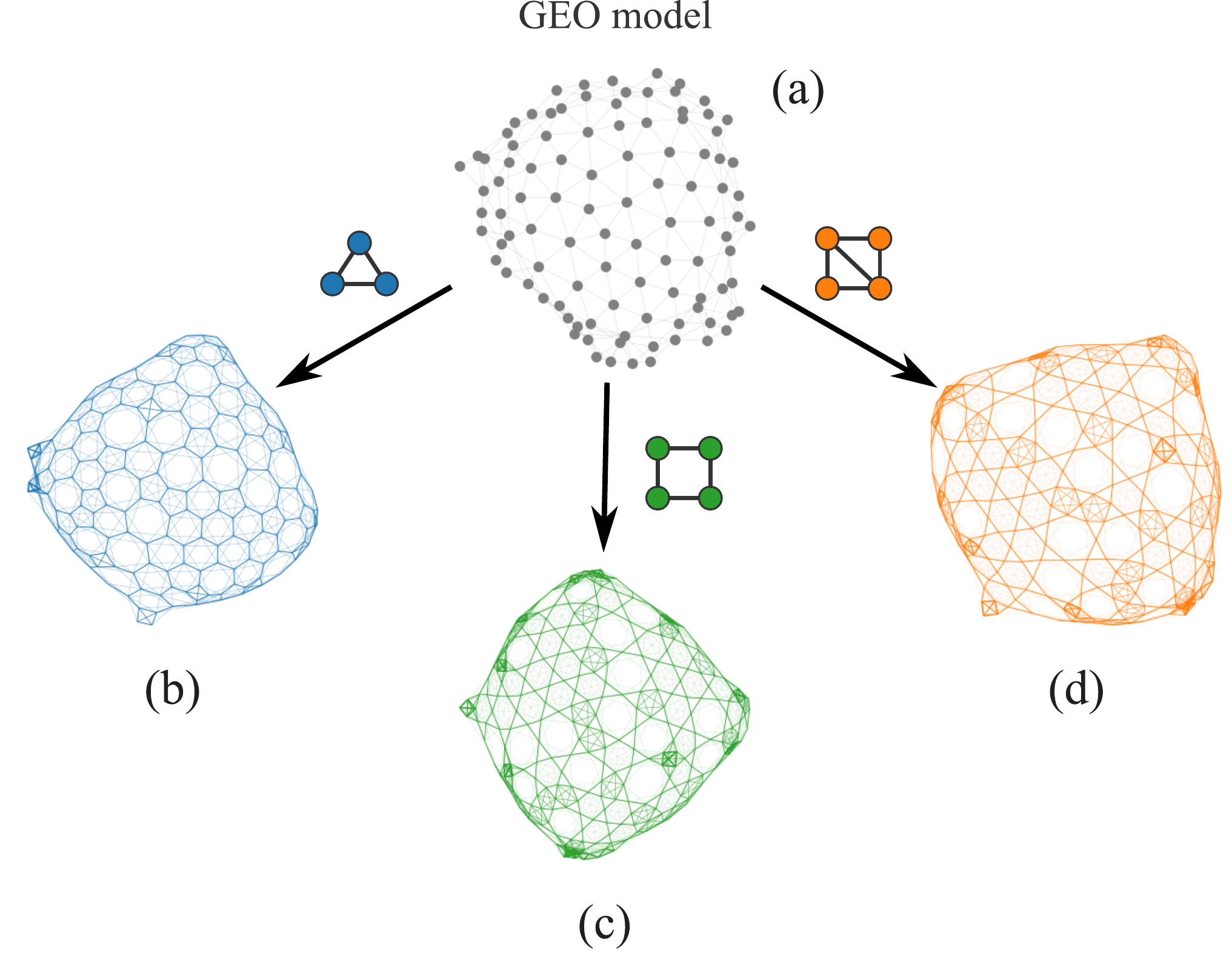}
   \caption{Illustration of a GEO network (with 100 nodes and $<k=5.72>$) (a) and the respective associated networks obtained for the triangle (b), square (c), and slashed square (d) agents.}\label{fig:net2equiv_GEO}
\end{figure}

Figures~\ref{fig:net2equiv_ER}--\ref{fig:net2equiv_GEO} show examples of network models (ER, BA, GEO) and respective associated networks derived for the three types of agents under consideration. The self-loops have been excluded from the visualizations of the associated networks, for simplicity's sake. The number of nodes, average degree and average strength of these networks are summarized in Table~\ref{tab:associated}.

\begin{table}[!ht]
\caption{Number of nodes ($N$), average degree ($<k>$), and average strength ($<s>$) for complex network models (ER, BA, GEO) and their respective associated networks derived for different considered agents. Observe that associated networks obtained for single-point agents are identical to the original networks.}
\label{tab:associated}
\begin{tabular}{|c|c|c|c|c|}
\hline

\begin{tabular}[c]{@{}c@{}}Network\\ Model\end{tabular} & Agent & $N$   & $<k>$ & $<s>$       \\ \hline \hline

\multirow{4}{*}{ER}  & single-point     & 100  & 12.00 $\pm$ 3.18     & 12.00 $\pm$ 3.18      \\
                     & triangle         & 293  & 17.24 $\pm$ 7.01     & 44.20 $\pm$ 14.71     \\
                     & square           & 2424 & 112.48 $\pm$ 37.33   & 241.57 $\pm$ 105.96   \\
                     & slashed square   & 578  & 38.74 $\pm$ 15.68    & 161.53 $\pm$ 59.45    \\ \hline
\multirow{4}{*}{BA}  & single-point     & 100  & 11.58 $\pm$ 7.42     & 11.58 $\pm$ 7.42      \\
                     & triangle         & 575  & 159.63 $\pm$ 92.03   & 352.98 $\pm$ 268.62   \\
                     & square           & 5468 & 1258.58 $\pm$ 716.81 & 3865.34 $\pm$ 3436.47 \\
                     & slashed square   & 2939 & 1018.45 $\pm$ 488.73 & 4052.20 $\pm$ 3127.55 \\ \hline
\multirow{4}{*}{GEO} & single-point     & 100  & 5.72 $\pm$ 1.12      & 5.72 $\pm$ 1.12       \\
                     & triangle         & 189  & 9.59 $\pm$ 1.12      & 30.50 $\pm$ 3.03       \\
                     & square           & 287  & 14.42 $\pm$ 1.81     & 82.03 $\pm$ 16.12     \\
                     & slashed square   & 276  & 13.71 $\pm$ 1.64     & 76.16 $\pm$ 12.80     \\ \hline
\end{tabular}
\end{table}

In the case of the ER networks shown in Figure~\ref{fig:net2equiv_ER}, substantially distinct associated networks have been obtained for each of the three agents. These networks, and especially that obtained for the square agent, are characterized by relatively dense interconnectivity as shown in Table~\ref{tab:associated}.  In general, among the three considered network models, the BA model tended to yield the densest associated networks. The associated networks obtained in the case of the GEO model (see Fig.~\ref{fig:net2equiv_GEO}) have similar properties (number of nodes, average degree, and average strength) according to Table~\ref{tab:associated}. 

It can be also observed from Figures~\ref{fig:net2equiv_ER}--\ref{fig:net2equiv_GEO} and Table~\ref{tab:associated}, that the square agent tended to imply denser associated networks. According to Table~\ref{tab:associated}, the associated networks obtained for the BA network in Figure~\ref{fig:net2equiv_BA} resulted even denser than those obtained for the ER and GEO models.

\begin{figure}[!ht]
  \centering
     \includegraphics[width=.8 \textwidth]{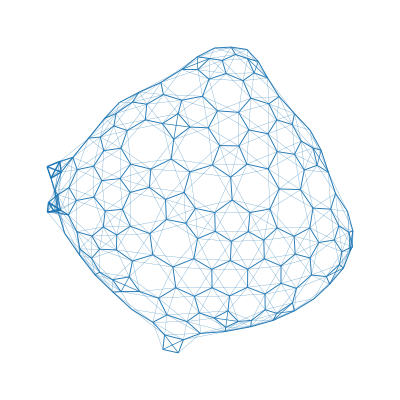}
   \caption{Magnification of the associated network in Fig.~\ref{fig:net2equiv_GEO}(b), obtained by considering the triangle agent on the original ER network in Fig.~\ref{fig:net2equiv_GEO}(a).}\label{fig:equvalentNet_GEO_0}
\end{figure}

Figure~\ref{fig:equvalentNet_GEO_0} shows, in more detail, an associated network obtained for the triangular agent from an ER network. This network corresponds to a magnification of the structure in Figure~\ref{fig:net2equiv_GEO}(b).

The tendency to geometric symmetry in this network can be understood as a consequence of the geometric and topological regularity of the original network (GEO). At the same time, the diversity of polygons observed in Figure~\ref{fig:equvalentNet_GEO_0} are mostly a reflection of the statistical fluctuations in the original network.

\subsection{Uniform Random Walks}\label{sec:random}

Thus far, we have addressed several structural aspects of the associated networks obtained for the considered models and agents.  Now, we proceed to study the dynamic properties of random walks performed uniformly in those types of associated networks.

\begin{figure}[!ht]
  \centering
     \includegraphics[width=.49 \textwidth]{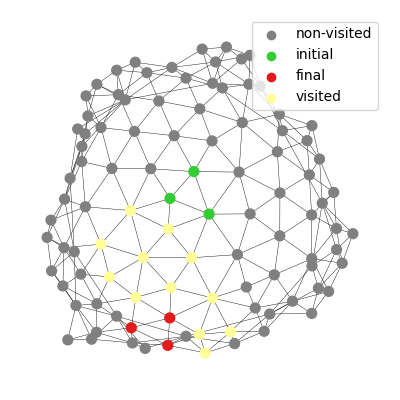}
     \includegraphics[width=.49 \textwidth]{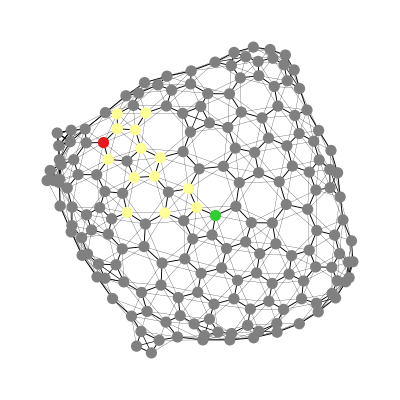} \\
     \vspace{-.3 cm}
     \hspace{.3 cm} (a) \hspace{6.5 cm} (b)
   \caption{Examples of a trajectory defined by 30 steps of a random walk performed by a triangle agent on a GEO network visualized in (a). The respective associated network is shown in (b). The network locus initially occupied by the agent is indicated in green, and its final locus is presented in red. The nodes covered along the 30 steps of the random walks are shown in yellow, while nodes that were not visited are depicted in gray.}\label{fig:walks}
\end{figure}

Figure~\ref{fig:walks} illustrates examples of \emph{trajectories} resulting from 30 steps of a random walk performed by a triangle agent on a GEO network (a), as well as the respective traditional uniform random walks resulting in the associated network (b). Observe that these two random walk realizations have the same dynamical properties.

In this work, for sake of computational efficiency, the results have been obtained by performing uniform random walks in the associated network.

\section{Methodology}

\begin{figure}[!ht]
  \centering
     \includegraphics[width=.85 \textwidth]{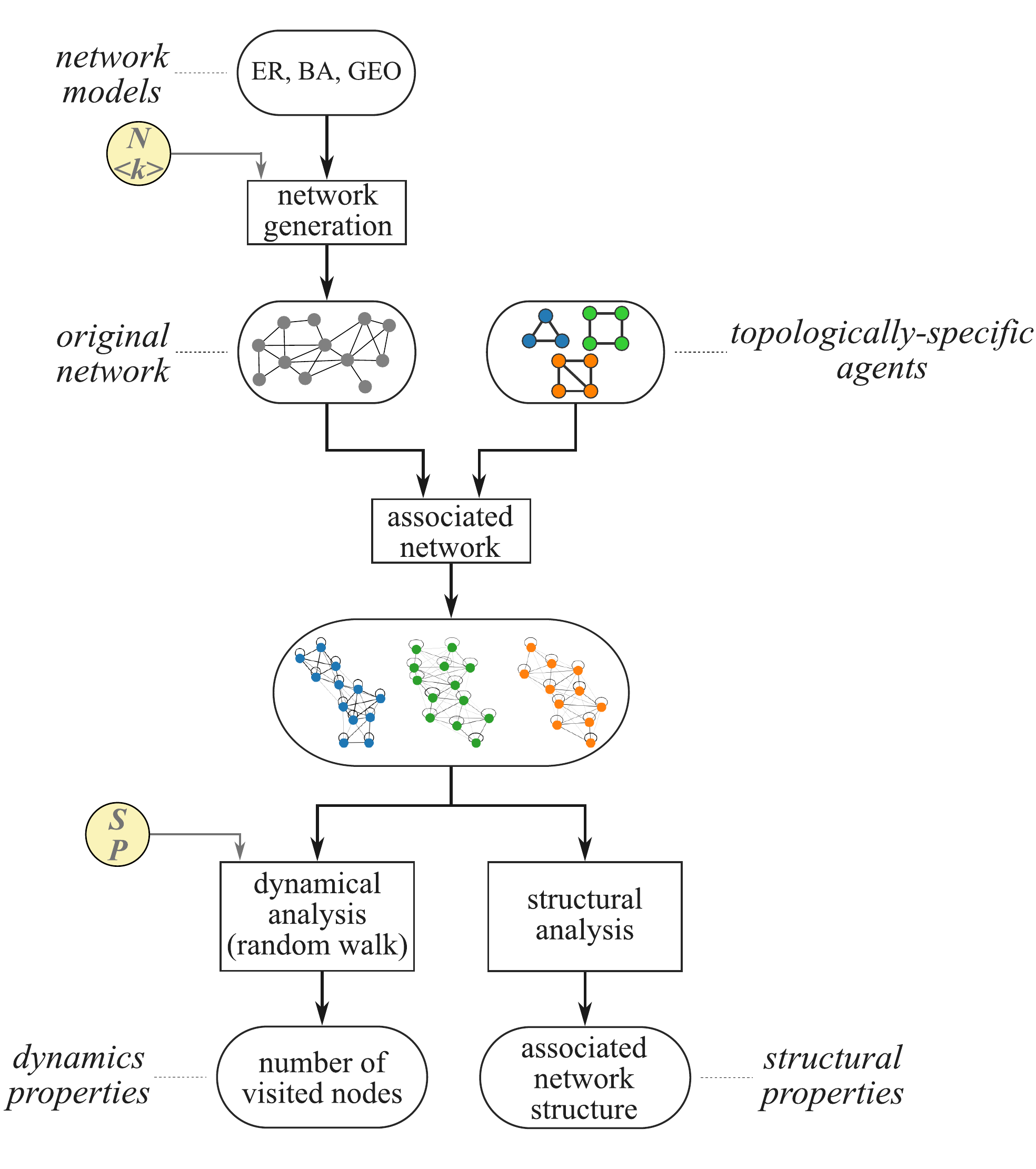}\\
   \caption{Flow diagram summarizing the main data (round corner boxes) and methods (square boxes) involved in the analysis performed in this work, while considering three network models and three topologically-specific agents. The parameters involved in some of the methods are indicated in yellow disks.}\label{fig:diagram}
\end{figure}

Figure~\ref{fig:diagram} illustrates the integration of the several concepts and methods adopted in the present work and discussed in the previous sections, which include the generation of the original network (Sec. \ref{sec:models}), followed by the calculation of the respective associated network (Sec. \ref{sec:equivNet}). Random walks are then performed on these networks, and their structural (network admissibility, connected component sizes, connected component number) and dynamic properties (number of visited nodes) are obtained.

The \emph{node coverage} of a given complex network by a specific agent is here quantified in terms of the average and standard deviation of the number of distinct nodes visited by the random walk after a given number of respective steps.

\section{Experimental Results}

In this section, we present, along subsequent subsections: (i) the structural analysis of the associated networks (ii) the dynamical analysis taking into account the node coverages obtained for uniformly random walks in GEO, ER, and BA networks as performed by the three types of topologically-specific agents as well as a single point agent (traditional random walk).

\subsection{Structural Analysis}\label{sec:Strutural}

Though the present work focuses interest on dynamical properties of random walks performed by topologically-specific agents, it is also of particular interest to develop a preliminary analysis of the considered associated networks in order to better understand their topological properties, as they can influence the respective random walk dynamics.

Three main measurements are considered henceforth: (i) network admissibility; (ii) number of connected components; (iii) sizes of the connected components. The \emph{network admissibility} corresponds to the percentage of nodes within the original network belonging to the positions where the topologically-specific agent fits. In order to provide an indication of how sparse each associated network is, the respective \emph{number of connected components} and \emph{their sizes} are also considered henceforth.  

\begin{figure}[!ht]
  \centering
     \includegraphics[width=.99 \textwidth]{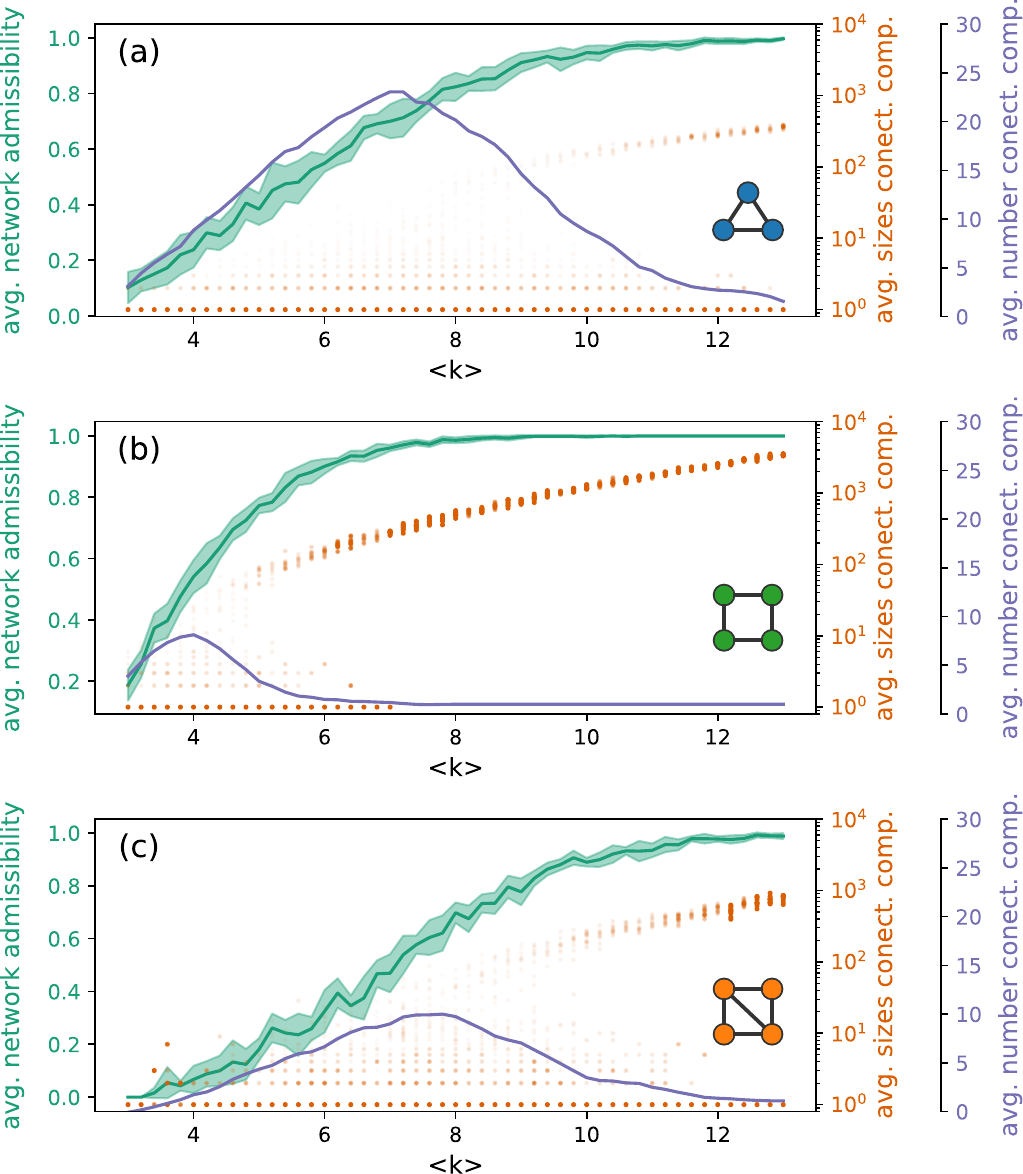}
\caption{The structure of the associated network obtained for ER model (for several values $\left<k \right>$ of average degree) respectively to the three considered agents: (a) triangle, (b) square, and (c) slashed square. The plots include the histogram of the average sizes of the connected components in the original network (in orange), the average number of connected components in the original network (violet), as well as the average network admissibility (in green). }
\label{fig:occupations_ER}
\end{figure}

Simultaneously presented in Figure~\ref{fig:occupations_ER}, for the case of the ER network model, are: (i) the average $\pm$ standard deviation of the network admissibility (in green); (ii) the average of the number of connected components (in blue); and (iii) a set of histograms of the connected components sizes (in orange).

The structural analysis in the ER model revealed that the admissibility (in green), represented by green curve, undergoes increase with saturation until the value 1.0 for the three considered cases, while a faster increase has been observed for the topologically-specific square agent (see Fig.~\ref{fig:occupations_ER}(b)).

In addition, the number of connected components, represented by the blue curve, exhibits a peak in all cases. The magnitude of this peak is larger for the triangle agent (see Fig.~\ref{fig:occupations_ER}(a)) and, in the case of the square agent, this peak occurs at a smaller value of $k$, also taking smaller intensity.

The histograms of connected components sizes (in orange) show that, for low values of $k$, the associated networks are composed of small components, the majority of them being of size 1 (i.e.~are isolated nodes). However, as $k$ increases, the number of components decreases as they merge, resulting in smaller but more substantial components.

For higher values of $k$, particularly around $k \approx 12$, the associated networks tended to reach a stabilized state for all cases under consideration (see Figs.~\ref{fig:occupations_ER}(a -- c)). More specifically, admissibility then reaches saturation, and the networks become fully connected.

\begin{figure}[!ht]
  \centering
     \includegraphics[width=.99 \textwidth]{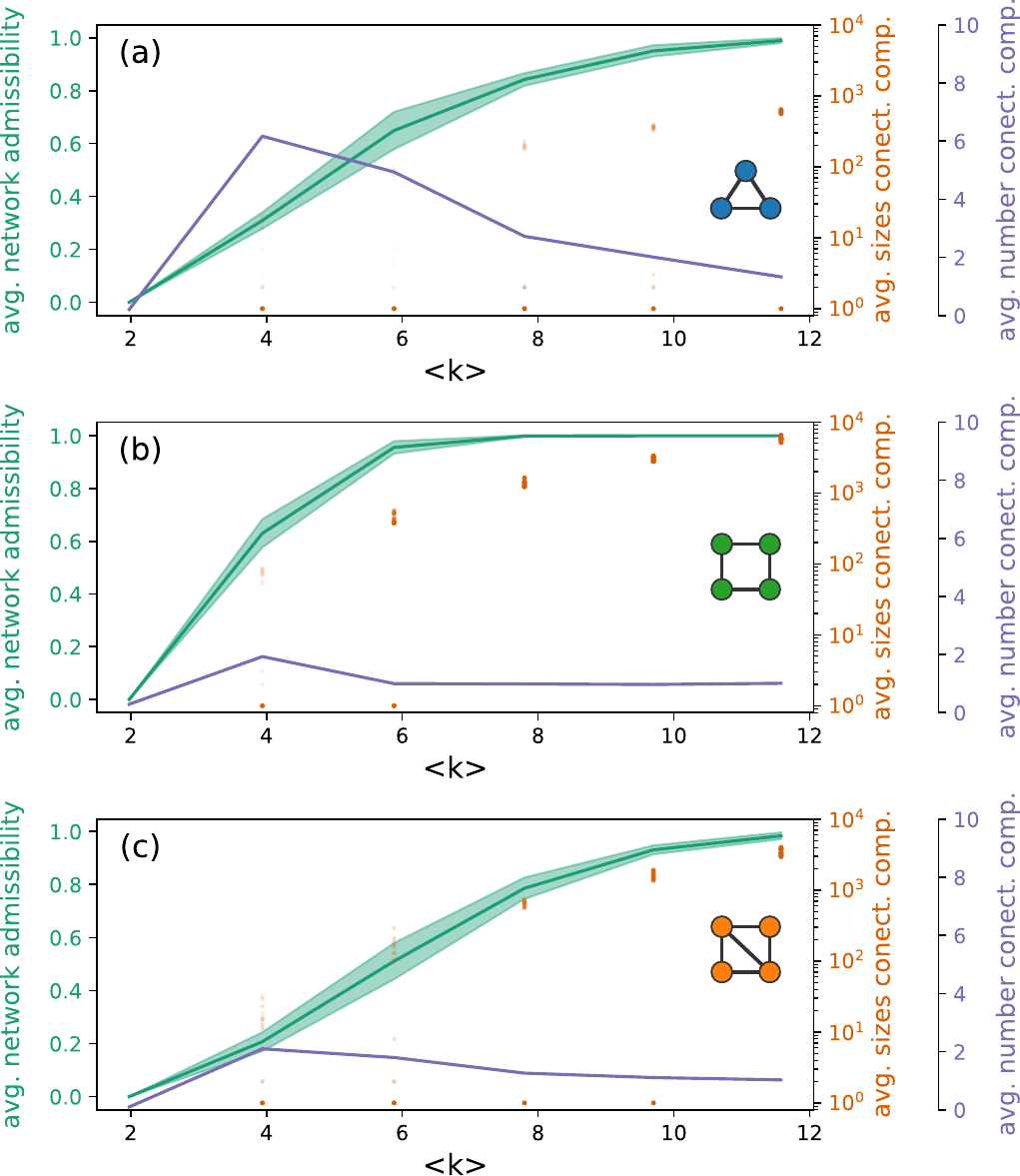}
\caption{The structure of the associated network obtained for BA model (for several values $\left<k \right>$ of average degree) respectively to the three considered agents: (a) triangle, (b) square, and (c) slashed square. The plots include the histogram of the average sizes of the connected components in the original network (in orange), the average number of connected components in the original network (violet), as well as the average network admissibility (in green). }
\label{fig:occupations_BA}
\end{figure}

In addition, the structural analyses for the BA network model were considered, and these are presented in Figure~\ref{fig:occupations_BA}.

The results from the structural analysis for the BA are analogous to those observed in the ER networks model. However, some differences can be highlighted. More specifically, the admissibility values have a slow increase when the slashed square is considered as topologically-specific agent (Fig.~\ref{fig:occupations_BA}(c)). In addition, the average number of connected components remains higher for the triangle agent (Fig.~\ref{fig:occupations_BA}(a)), but now the peak occurs for smaller values of $k$.

Note that no diagrams similar to those in Figures~\ref{fig:occupations_ER} and~\ref{fig:occupations_BA} have been obtained for the GEO model because networks of this type, having the same number of nodes have the same average degree ($k$) values, which are determined by the spatial distribution of points. For any GEO network, the associated network invariably yields a single component and an admissibility value of 1.0 for all topologically-specific agents considered.

\subsection{Node Coverage by Uniformly Random Walks}

The second part of the current study addresses the dynamic analysis of random walks performed by topologically-specific agents, focusing on respective node coverage.

In this study, three network models were considered: ER, BA, and GEO.
It is important to note that all original networks were generated with 100 nodes, and the choice of the average degree ($k$) value to be adopted for generating the networks was based on the structural analysis described in Section~\ref{sec:Strutural}, aimed at obtaining associated networks having a single connected component. Furthermore, additional experiments were performed considering the same original network without the five largest degree nodes.

\begin{figure}[!ht]
  \centering
     \includegraphics[width=.99 \textwidth]{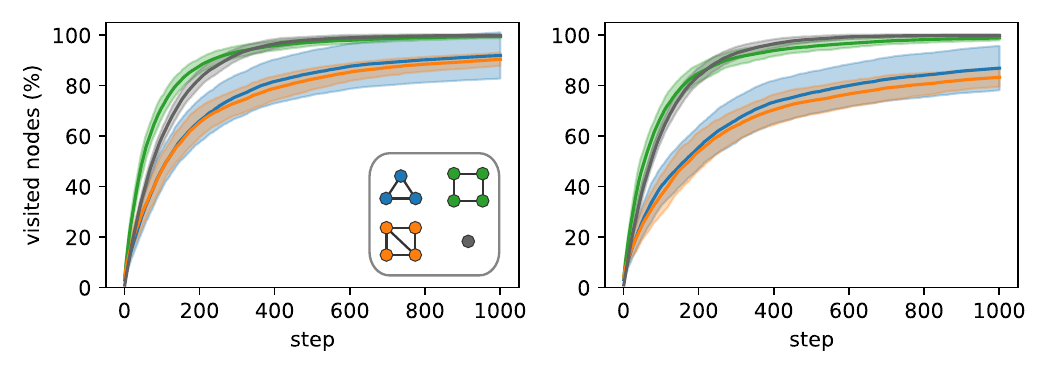}\\
     \hspace{.7 cm} (a) \hspace{5.6 cm} (b)
\caption{Percentage of visited nodes in terms of performed steps considering four agents (triangle, square, slashed square, and point): (a) over an ER model with $N = 100$ and $<k> = 12$; and (b) the same network after removing the five nodes with the largest degrees (hubs).}\label{fig:randomWalk_ER}
\end{figure}

\begin{figure}[!ht]
  \centering
     \includegraphics[width=.99 \textwidth]{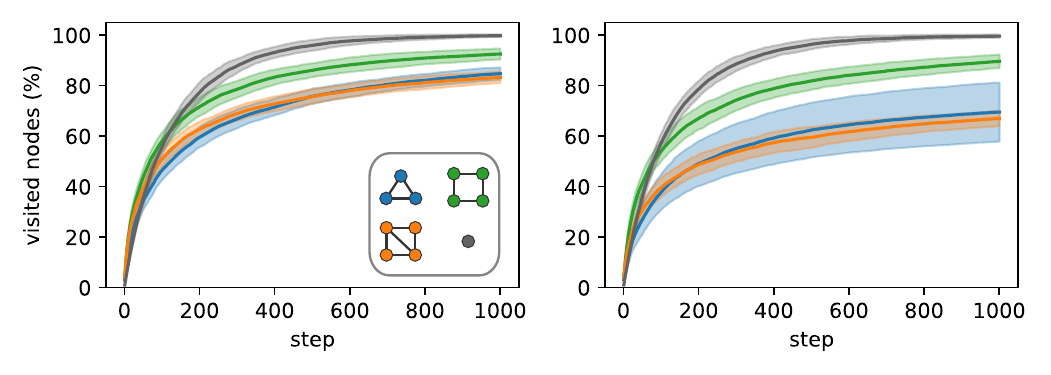}\\
     \hspace{.7 cm} (a) \hspace{5.6 cm} (b)
\caption{Percentage of visited nodes in terms of performed steps considering four agents (triangle, square, slashed square, and point): (a) over a BA model with $N = 100$ and $<k> = 11.58$; and (b) the same network after removing the five nodes with the largest degrees (hubs).}\label{fig:randomWalk_BA}
\end{figure}

\begin{figure}[!ht]
  \centering
     \includegraphics[width=.99 \textwidth]{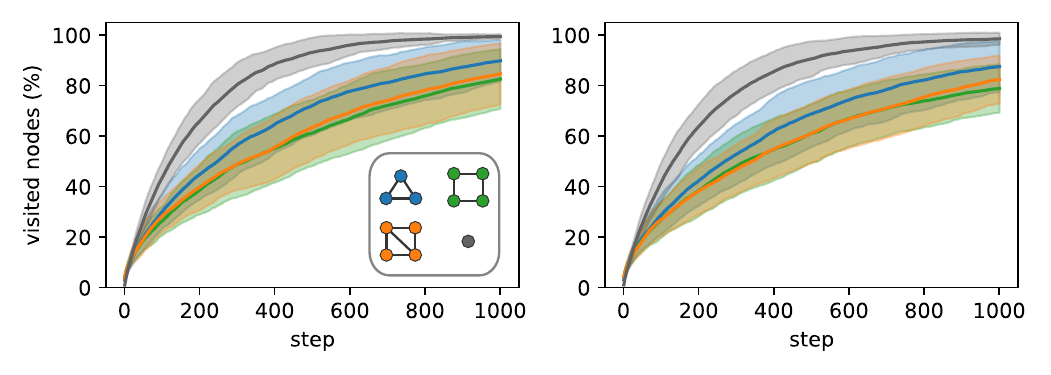}\\
     \hspace{.7 cm} (a) \hspace{5.6 cm} (b)
\caption{Percentage of visited nodes in terms of performed steps considering four agents (triangle, square, slashed square, and point): (a) over a GEO model with $N = 100$ and $<k> = 5.72$; and (b) the same network after removing the five nodes with the largest degrees (hubs).}\label{fig:randomWalk_GEO}
\end{figure}

The random walk analysis was performed 100 times on the associated networks of the ER (Fig.~\ref{fig:randomWalk_ER}), BA (Fig.~\ref{fig:randomWalk_BA}), and GEO (Fig.~\ref{fig:randomWalk_GEO}) models respectively to three topologically-specific agents: triangle, square, and slashed square. A conventional random walk (single point) was also included for comparative purposes.

Several results can be observed. First, we have that the triangle agent exhibited the greatest dispersion, with the GEO network model showing higher levels of dispersion for all agents analyzed. In the ER model, the square agent yielded coverage behavior similar to the traditional single-point random walk in both scenarios (original network and the same with its five largest hubs removed), being the fastest in covering the network. Interestingly, the triangle and slashed square agents yielded similar dynamical features for the ER model.

The removal of five nodes did not cause significant changes in the GEO model. However, in the ER model, the performance of the triangle and slashed square agents to cover the network were undermined, and, in the case of BA model, the impact was more pronounced, with the triangle and slashed square agents moving more slowly, while the triangle underwent substantially greater dispersion. In the GEO model, the square agent yielded dynamics similar to that of the slashed square agent in both scenarios, while the triangle agent tended to yielde the fastest network coverage, after the single-point random walk. Among the network models analyzed, the GEO case generally led to the slowest agent coverage.

All in all, the conventional random walk executed by a single point tended to be the most efficacious approach for covering the network in most considered cases. In contrast, the slashed square agent tended to exhibit the slowest network coverage.

\section{Concluding Remarks}

While the areas of random walks and complex networks have provided the subject for several studies, their combination (i.e.~random walks in complex networks) motivates particular potential interest as a perspective from which to integrate studies of the interrelationship between the network topology and the dynamics of the respective random walks. However, most approaches to random walks on complex networks have considered agents that correspond to a single node, which can therefore fit into any of the nodes of the respectively explored network.

In the present work, concepts and methods allowing random walks involving moving agents that have specific topology have been described and considered for respective studies concerning three basic types of topologically-specific agents (triangle, square, and slashed square) and three complex networks (GEO, ER, and BA).

Basically, the agents are taken to correspond to relatively small graphs, which need to fit the local topology of the networks in which they are to perform random walks. More specifically, an agent has been understood to fit a specific locus of the network provided that locus is a subgraph that contains the agent's graph. The displacement of these types of agents was then assumed to take place between adjacent admissible sites of the network. By adjacent, it means that each of the nodes in the new position are zero or first neighbors of the respective node of the previous position.

Given a topologically-specific agent and a network to be covered, it is possible to obtain the respective \emph{associated networks}, which corresponds to a traditional network whose each node represents a possible position of the agent in the original network, while the links indicate possible displacements between those positions (see Sec.~\ref{sec:equivNet}).

Several interesting results have been reported and discussed in the present work, being summarized as follows respectively to the two main covered topics, namely structural analysis of the associated networks and characterization of the respectively obtained random walk dynamics.

We analyzed three key structural aspects of associated networks to the degree of the original network: (i) network admissibility, (ii) the number of connected components, and (iii) the size of these components. These metrics were examined for ER and BA network models.

It has been verified that the associated network attains a maximum number of components at a particular value of average degree. This maximum has been observed to be higher for the triangle agent. Additionally, as the mean degree exceeds a certain threshold, the number of components decreases as their size increases, suggesting a transition toward increased interconnectivity. Furthermore, for higher values of the average degree, particularly around 12, the associated networks showed a tendency to stabilize, presenting consistent characteristics across the considered cases.

Two properties of the random walks have been considered in the reported studies: (i) average node coverage; and (ii) standard deviation of the node coverage. Regarding the coverage of nodes allowed by the performed random walks, we observed that the random walk by using a single-point agent exhibited the best performance in covering the original network compared to the other agents.

Within the group of topologically-specific agents, the square yielded the most effective coverage in the ER and BA network models. In the context of the GEO network, the triangle agent led to slightly better results. Regarding the standard deviations of the node coverage, the triangle agent presented the respective highest values. 

Also of particular interest was the verification that substantially larger standard deviations and smallest node coverage efficiency have been obtained for the GEO networks, as compared to the other two topologies. This is possibly because the GEO structure presents an intrinsic local heterogeneity, implying the agent to be delayed at specific network positions, therefore leading to respective plateaus along the respective node coverage curves.

Regarding the experiment involving the hub removals, it has been verified that this type of network modification had relatively little effect for all considered agents and network models.

The described concepts, methods, and results motivate several further developments. As a matter of fact, any previously addressed aspect of random walks on complex networks can be revisited from the perspective of the currently suggested approach. However, the adoption of agents presenting specific topology also motivates possible further research aimed at relating the properties of specific types of networks and specific topologically-specific agents.

Prospects for further research include, but are not limited to the following cases. It would be interesting to consider more types of agents and networks, as well as other types of motion involving preferential displacements, as well as presence of external fields and other influences.  Another interesting perspective would be to consider the interaction (e.g.~in terms of fields) between more than one agent within the same or distinct topologies while moving within a same complex network. Also of interest would be to consider agents that can adapt (morph) their topology as a consequence of reaching particular situations (e.g.~getting stuck) or even periodically.

\section*{Acknowledgments}
Alexandre Benatti thanks MCTI PPI-SOFTEX (TIC 13 DOU 01245.0102\\22/2022-44).
Luciano da F. Costa thanks CNPq (grant no.~313505/2023-3) and FAPESP (grants 15/22308-2 and 2022/15304-4).

\bibliography{ref}
\bibliographystyle{unsrt}

\end{document}